# Visualizing Crystallization Dynamics and Transformation Pathways of Disordered Rocksalt Oxides During Thermally Activated Sol-Gel Synthesis

Diyi Cheng[1,§], Tim Kodalle[2,3,§], Anika T. Promi[2], Ansuman Halder[2], Raphael F. Moral[2,4], Madeline Grass[2,5], Venkata S. Avvaru[1], Haegyeom Kim[1], Carolin M. Sutter-Fella[2,*], Haimei Zheng[1,6,*]

**Affiliations:**
[1]Materials Sciences Division, Lawrence Berkeley National Laboratory, Berkeley, CA 94720, USA
[2]Molecular Foundry, Lawrence Berkeley National Laboratory, Berkeley, CA 94720, USA
[3]Advanced Light Source, Lawrence Berkeley National Laboratory, Berkeley, CA 94720, USA
[4]Brazilian Synchrotron Light Laboratory (LNLS), Brazilian Center for Research in Energy and Materials (CNPEM), Campinas, Sao Paulo 13083-970, Brazil
[5]Department of Chemical and Biomolecular Engineering, University of California, Berkeley, California 94720, USA
[6]Department of Materials Science and Engineering, University of California, Berkeley, CA 94720, USA

*Corresponding author E-mails: hmzheng@lbl.gov (H. Zheng), csutterfella@lbl.gov (C.M. Sutter-Fella)
§These authors contributed equally

**Abstract:**

Sol-gel synthesis is a wet-chemical processing route for the fabrication of functional materials offering control over composition, morphology, and microstructure at relatively low processing temperatures compared to conventional solid-state synthesis methods. While the sol-gel process initiates with intermixed molecular precursors, the transformation pathways at the early nucleation stage are insufficiently understood. Here, the chemical and structural transformation of disordered rocksalt (DRX) $Li_{1.2}Mn_{0.4}Ti_{0.4}O_2$ (LMTO), a promising cathode material for lithium batteries, is studied by multiscale characterization tools. *In situ* heating transmission electron microscopy (TEM) using a liquid cell visualizes and identifies crystallization pathways at nanoscale. While some regions follow a classical multi-step transition through thermodynamically stable intermediates, others exhibit a kinetic shortcut in which intermediate nanocrystals dissolve into a localized amorphous matrix that directly precipitates the DRX structure. Macroscale FTIR corroborates the findings to be related to chemically distinct microenvironments in the gel precursor, with transition metal ions more strongly incorporated into the acetate-coordinated network than lithium. Although *in situ* heating TEM captures diverse local transformation pathways, *in situ* SXRD indicates that the macroscopic transformation proceeds predominantly through spinel LMTO and lithium titanite intermediates toward DRX-LMTO. The findings shed light on the spatiotemporal chemical and structural transformations in sol-gel derived DRX-LMTO materials, and call for fine tuning of such sol-gel chemistries to manipulate the crystallization pathways and achieve target material homogeneity more efficiently.

**Introduction**

The sol-gel process has long been established as a versatile, bottom-up wet-chemical technique for synthesizing nanostructured metal oxides, glasses, and functional ceramics[1,2]. Through concurrent hydrolysis and polycondensation of molecular precursors, typically metal alkoxides or metal salts dissolved in aqueous or organic solvents, this process bypasses the limitations of traditional solid-state synthesis by enabling the mixing of constituent elements at the molecular level[3,4]. State-of-the-art advancements in sol-gel chemistry have demonstrated that the kinetics of these fundamental reactions can be precisely tuned[5–7]. Variables such as solvent polarity, pH control, and the introduction of complexing or chelating agents dictate the structural evolution of the material from its early stages[8,9]. One good example is the choice of solvent that could profoundly influence precursor dissolution and the homogeneity of the resulting colloidal network. Nonaqueous or aprotic solvents tend to coordinate effectively with transition metal ions, mitigating hyperactive hydrolysis to form homogeneous, clear sols[10]. Conversely, protic solvents often lead to rapid, uncontrolled hydrolysis, triggering localized precipitation and nanoscale particle segregation[3,4]. Despite these macroscopic kinetic controls, the degree of spatial heterogeneity within the gel precursor, and how it governs the nucleation of transient intermediates during evolution toward the target phase remains unclear.

Insight from these underlying nucleation and subsequent growth mechanisms is particularly essential in the development of functional materials. The model system investigated here, disordered rock salt (DRX), is of technological importance due to globally increasing demand for high-performance lithium batteries[11–14]. DRX oxides have emerged as promising, scalable cathode candidates owing to the enhanced capacity from excess lithium content, three-dimentional lithium diffusion pathways, and reduced mechanical strain accumulation[15–17], etc. Specifically, manganese- and titanium-based DRX materials, i.e., $Li_{1.2}Mn_{0.4}Ti_{0.4}O_2$ (LMTO), offer improved specific capacities utilizing purely earth-abundant elements[15]. However, in order to improve lithium diffusion kinetics, synthesizing phase-pure, small-grained DRX-LMTO has proven challenging, primarily due to the high-temperature processing via traditional solid-state synthesis routines[16,18]. As such, sol-gel synthesis presents an attractive alternative capable of lowering the kinetic barriers and potentially synthesis temperature by mixing transition metal species at the molecular level prior to thermal treatment. Nevertheless, sol-gel process introduces its own set of complex, unresolved issues during the critical gel-to-crystalline transformation stage[19]. One of the major hurdles lies in managing the differential reactivity of the multi-metallic precursors. Titanium precursors, such as titanium isopropoxide, are sensitive to moisture and tend to undergo rapid hydrolysis[9,19]. If the solvent system does not suppress this reactivity, the titanium species will precipitate out as localized titanium-rich clusters before the manganese precursors can fully integrate into the polymer network[19]. This nanoscale chemical segregation within the seemingly uniform amorphous xerogel hinders homogeneous nucleation and influences the crystallization pathways[20]. Consequently, during intermediate drying and pyrolysis stages (typically occurring at up to 400 °C), the localized concentration heterogeneities lead to a variety of competing, thermodynamically stable intermediate phases, such as spinel-LMTO (S-LMTO), lithium titanites (LTO), lithium manganese oxides (LMO), and lithium-free manganese titanites (MTO)[18,19].

Regulating these heterogeneous crystallization events requires the understanding of nanoscale compositional microenvironments to finetune the initial precursor chemistry and thus tailor the transformation pathway, a task for which conventional characterization techniques, such as X-ray diffraction (XRD), are not ideally suited. *In situ* liquid cell transmission electron microscopy (TEM), with the capability of capturing material dynamics down to the atomic level, offers a new paradigm to resolve such unexplored, complex regimes during sol-gel synthesis[21,22]. By utilizing liquid cell configuration and heating capabilities, one has the chance to access nanoscale fluid behaviors, observe the initial collision, hydrolysis, and agglomeration of the metal precursors in their native liquid solvent or gel environments[21,23]. Such observations would enable the direct mapping of how specific solvent polarities dictate the spatial distribution of early colloidal networks[24]. One can also continuously monitor localized atomic migrations, directly identifying whether nucleation begins homogeneously within the bulk of the gel or heterogeneously at the surface, while simultaneously tracking the sequence of intermediate phase formation that leads to the final desired structure[24,25].

However, employing *in situ* liquid cell TEM to probe through the sol-gel evolution, from wet liquid precursors to calcined solid particles, presents several technical challenges[26]. Liquid cell TEM relies on encapsulating the solvent between ultra-thin, electron-transparent membranes to maintain a liquid state within the high-vacuum environment of the microscope[23]. The enclosed liquid cells can be susceptible to the high temperatures required for calcination, where the applied heat increases the vapor pressure of the solvent and may rupture the thin membranes at elevated temperatures[21,26]. Besides, the complex nucleation events and crystallization dynamics at the early stage of pyrolysis, i.e., below 400 °C, cast tremendous difficulties on data analysis and interpretation that can be easily biased via manual processing. Therefore, previous studies have predominantly focused on material transformations at higher temperatures, where the early nucleation subsides, nanocrystals appear larger and the subsequent crystallization becomes less convoluted[27–30]. Additionally, the electron beam inherently interacts with the precursors and remaining solvents, which may artificially alter the nucleation pathways unless carefully controlled[26,31].

In this work, we leveraged *in situ* liquid cell TEM with heating capability to investigate the crystallization dynamics of DRX-LMTO through the early stage of pyrolysis (25-400 °C) in a liquid cell to a later stage of calcination in vacuum. The xerogel state of samples prevents excess solvent evaporation during heating. A customized suite of algorithms was developed based on virtual dark-field (VDF) analysis, automating precise phase identification, trackable phase mapping, and evolutionary phase transformation overlaying real-space images. By carefully suppressing beam effects and thermal drifting, we captured the complex nucleation behaviors and identified unprecedented crystallization pathways during the pyrolysis and calcination processes of DRX-LMTO synthesis. At the initial nucleation stage (25-400 °C), binary intermediates (manganese/titanium oxides) can easily merge with ternary ones to form more favored S-LMTO phase, while ternary intermediates have difficulty merging with ternary analogues unless a critically small grain size is met. At the calcination stage (400-800 °C), the direct nucleation of the DRX-LMTO structure from amorphous phase was observed, bypassing the more thermodynamically favored S-LMTO intermediate phase. Along with macroscopic characterizations such as Fourier-transform infrared (FTIR) spectroscopy, dynamic light scattering (DLS), and *in situ* synchrotron XRD (SXRD), we provide a more complete view of materials transformations, both chemically and temporospatially, during mixing, gelation, nucleation, growth and eventual phase formation of thermally activated sol-gel synthesis. Discovered unexpected nanoscale phase transformation provides new insights on how the distinct microenvironments within a sol-gel derived precursor influences the crystallization dynamics of DRX-LMTO material, which could be potentially utilized to circumvent the high-temperature requirement during synthesis.

**Results and Discussion**

**Figure 1a** outlines the multiscale experimental workflow, utilizing macroscopic spectroscopy alongside electron microscopy and X-ray characterization to track material transformations and dynamics. From liquid precursors to calcined solid DRX-LMTO powder, FTIR and DLS are first utilized to examine the chemical signature and colloidal size distribution of the precursor solution; Scanning transmission electron microscopy (STEM) and electron energy dispersive spectroscopy (EDS) are then used to examine the gel after solvent evaporation and network consolidation; *In situ* liquid cell heating TEM and *in situ* SXRD are subsequently used to capture the transient structural transformation at nanoscale and macroscopic scale, respectively, along with morphological examination via scanning electron microscopy (SEM).

FTIR spectra in **Figure 1b** track the evolution from precursor solution (LMTO_MeOH_Sol) to gel (LMTO_MeOH_Gel), and reveal substantial changes in coordination environment and intermolecular interactions during the gelation process. In the precursor solution, distinct bands including $COO^-$ deformation modes (~600–800 cm$^{-1}$), scissoring modes (~900–1000 cm$^{-1}$), and a well-defined asymmetric carboxylate stretch near ~1560 cm$^{-1}$ are observed[32]. Examination of the individual metal-containing solutions confirms that this carboxylate signal arises predominantly from Mn–acetate and Ti–acetate coordination, while the Li-containing solution exhibits minimal contribution (**Figure S1**). The gel exhibits a pronounced increase in band broadening across the entire 600–1600 cm$^{-1}$ region, accompanied by a substantial reduction in transmittance in certain wavenumber regions, indicating the formation of a chemically heterogeneous and more strongly interacting network. The asymmetric COO⁻ stretch near ~1560 cm⁻¹ increases in intensity and broadens, indicating both a greater proportion of acetate ligands engaged in metal coordination and the emergence of diverse local coordination microenvironments, including bridging and chelating modes across multiple metal centers[32]. This is consistent with acetate acting as a cross-linking ligand within the developing network. Simultaneously, the lower-wavenumber region (~1300–1450 cm⁻¹) evolves into a wide, merged band, reflecting a transition from discrete molecular acetate environments to a distribution of metal–carboxylate coordination modes arising from interactions with multiple metal centers. The overall spectral homogenization exhibited by broad, overlapping bands replacing sharp solution-state features indicates that transition metal ions are no longer present as independent solvated species but are co-localized within a common coordination environment. While FTIR does not directly confirm heterometallic linkages, the observed redistribution, intensification, and broadening of carboxylate vibrations support the formation of a chemically integrated hybrid organic–inorganic gel network.

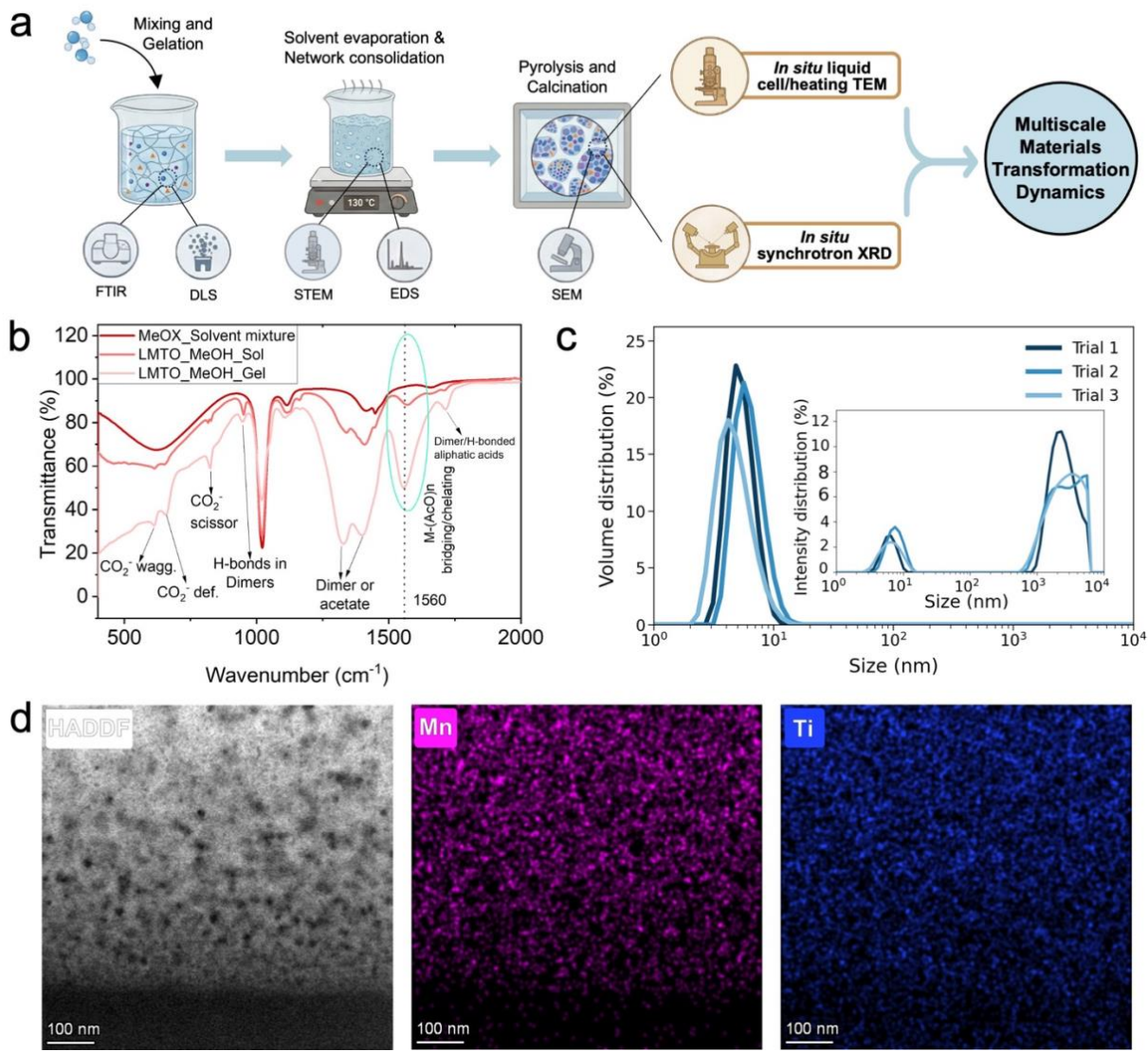

**Figure 1. Sol-gel synthesis workflow and sol/gel properties before pyrolysis**. **a**, schematic overview of the sol-gel synthesis process and the characterization tools used at each step, including FTIR, DLS, STEM-EDS, SEM, *in situ* liquid cell heating TEM, and *in situ* SXRD, for probing multiscale material transformation and crystallization dynamics. Note that gelation was performed at 70 °C. Pyrolysis was performed in the range of 130-400 °C in air and calcination was performed in the range of 400-1000 °C in argon atmosphere. **b**, FTIR spectra showing the evolution from precursor solution to gel. MeOX_Solvent Mixture represents the solvent mixture (methanol, water and acetic acid), LMTO_MeOH_Sol corresponds to the mixed precursor solution containing $LiNO_3$, $Mn(OAc)_2$, and $Ti(OiPr)_4$ precursor in methanol, acetic acid and water mixture. LMTO_MeOH_Gel denotes the gel obtained after gelation. **c**, DLS volume-weighted size distributions of LMTO precursor species in methanol. The solution contains lithium nitrate, manganese(II) acetate tetrahydrate, titanium(IV) isopropoxide, acetic acid, and methanol. Inset is the DLS intensity-weighted size distribution of LMTO precursor species in methanol. **d**, HADDF image and corresponding EDS mapping results of Mn and Ti elements of the same area on the gel, indicating a uniform elemental distribution of the precursors.

**Figure 1c** presents DLS results that are used to probe the size distribution of species in the LMTO precursor solution. The volume-weighted size distributions from three replicate measurements show a narrow, reproducible population centered at approximately 4–6 nm, indicating that the solution is dominated by small, dispersed metal–ligand clusters rather than large aggregates. The close overlap among replicates confirms good measurement consistency. The volume distribution is reported here in preference to the intensity distribution (**Figure 1c,** inset), as larger species scatter disproportionately more light and can obscure the true population composition; the intensity distribution does show a secondary peak at larger hydrodynamic sizes, attributable to a minor fraction of aggregated species formed upon addition of titanium(IV) isopropoxide and acetic acid, but this feature is not representative of the bulk solution. Together, these results indicate that the precursor solution before gelation consists predominantly of small nanoscale clusters in the few-nanometer range.

The gel sample was then examined by *ex situ* TEM after drying at 130 °C. The gel loaded onto nickel TEM grid appears fairly viscous (**Figure S2a**) and thinner sample region can be clearly seen from the low-magnification TEM image (**Figure S2b**). **Figure 1d** corroborates the general uniformity from DLS result through high-angle annular dark-field (HAADF) imaging paired with energy-dispersive X-ray spectroscopy (EDS) mappings under STEM mode. The STEM/EDS maps reveal a relatively homogeneous distribution of both manganese (Mn) and titanium (Ti) across the field of view, with no obvious agglomeration or elemental separation. The "island-like" features in the EDS maps are likely a result of local thickness differences, caused by the evaporation of residue solvents under the electron beam.

The data presented in **Figure 1** suggests that this specific methanol-based sol-gel route generates locally different microenvironments for transition metal species and lithium species at the gel state, despite an overall uniform distribution of Mn-/Ti-species. FTIR spectra show that the transition metals are chemically integrated into the polymeric gel network through robust metal-acetate coordination. STEM/EDS demonstrates the intermixing Mn- and Ti-species across the gel that could serve as the reservoirs of Mn and Ti ions for various intermediate phases during the subsequent heterogeneous nucleation events .

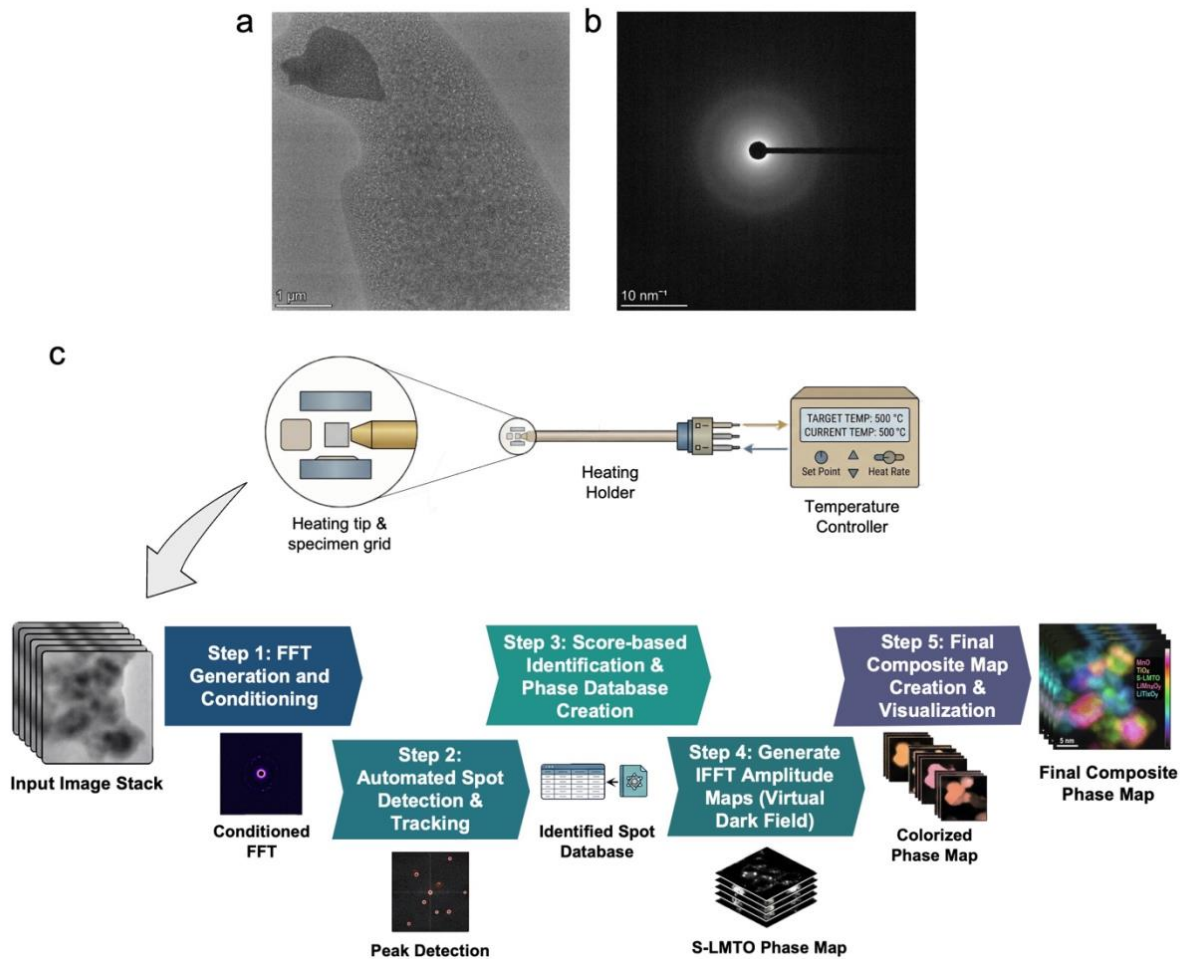

**Figure 2**. **Gel examination prior to *in situ* TEM measurement and the workflow for *in situ* dataset analysis**. **a** & **b**, Low-magnification morphology of LMTO gel before heating and corresponding SAED pattern showing the amorphous feature. **c**, *in situ* heating setup for the liquid cell configuration for the gel sample, along with the five-step virtual dark-field image analysis employed to generate the *in situ* datasets shown in Figure 3.

Prior to performing *in situ* heating TEM measurement, we first examined the beam stability and crystallinity of the LMTO gel. **Movie S1** demonstrates the beam stability of LMTO gel before heating, where continuous beam exposure does not promote nucleation or phase change within the gel except for some bubble generation, likely caused by the residue solvents in the gel. **Figure 2a&b** present a bright-field TEM image and the corresponding selected area electron diffraction (SAED) pattern of the gel sample, respectively. The relatively featureless image with uniform contrast in **Figure 2a** and the broad, diffuse halo in **Figure 2b** suggest the amorphous nature of the initial xerogel network prior to thermal heating beyond 130 °C. This observation fits to the FTIR findings confirming that transition metal ions are embedded in an amorphous gel network.

Next, **Figure 2c** details the integrated experimental and computational workflow for *in situ* heating TEM data collection and analysis. The sample was sandwiched between two nickel-based TEM grids as a common liquid cell configuration for high-temperature tolerance, and loaded onto a Gatan *in situ* heating holder (**Figure S3**). The *in situ* heating holder is utilized to precisely control the temperature while the microscope continuously captures high-resolution TEM (HRTEM) image stacks during material transformations. Note that manual HRTEM analysis struggles to deconvolute the simultaneous emergence of competing, highly localized nanophases that are buried within a dense, continuously evolving amorphous matrix. Therefore, it is of critical importance to develop proper, automated processing method in order to precisely identify the transient nucleation events and map out the structural transformations of the LMTO gel during heating.

As such, a custom five-step automated image processing pipeline was designed and shown in the schematic at the lower part of **Figure 2c**. The automated analysis workflow integrates principles of digital Fourier filtering and virtual dark-field (VDF) imaging to systematically identify and map nanoscale crystalline phases. In step 1, raw TEM frames undergo edge apodization and mean subtraction to suppress edge artifacts in the discrete Fast Fourier Transform (FFT) patterns. The algorithm then isolates distinct crystallographic reflections by subtracting a Gaussian-blurred background and detecting local intensity maxima through center-of-mass refinement and nearest-neighbor logic[33] as demonstrated in **Figure S4**. Following extraction, phase identification in step 3 is executed via a multi-metric scoring algorithm. This step dynamically scores each detected FFT spot against a predefined, system-specific crystallography dictionary using a weighted penalty matrix based on measured physical d-spacings, the theoretical XRD intensity rank for probable crystal planes, and expected temporal sequence of different phases. (**Table S1**) Finally in step 5, the workflow reconstructs real-space distribution maps of the identified phases by utilizing dynamic Gaussian masks and inverse Fast Fourier Transforms (IFFT) that extracts the local amplitude data to build individual phase maps[34,35]. By employing gamma-corrected alpha blending, the phase maps are color-coded and overlayed directly onto the real-space HRTEM image stacks, yielding a comprehensive, denoised visualization of structural evolution and phase propagation over time.

With the established workflow above, *in situ* heating TEM results were then collected, analyzed, and are shown in **Figure 3 & 4**. The heating sequence was divided into 25-400 °C for the pyrolysis stage, and 400-800 °C for the calcination stage, to match the temperature ranges where major phase transformations occur based on literature[19]. We pre-processed *in situ* video datasets using ImageJ software to account for thermal drifting issues, and facilitate subsequent automated image analysis. Details can be found in the Method section. **Figure 3** demonstrates the analyzed HRTEM image sequences from *in situ* heating TEM measurements up to 400 °C. The *in situ* video (**Movie S2**) captures three distinct, highly localized crystallization trajectories during the thermal activation of the LMTO gel. The 5-nm scale bars emphasize the nanoscale nature of the observations. By overlaying color-coded VDF maps onto the bright-field images, specific intermediate nanocrystals are dynamically tracked.

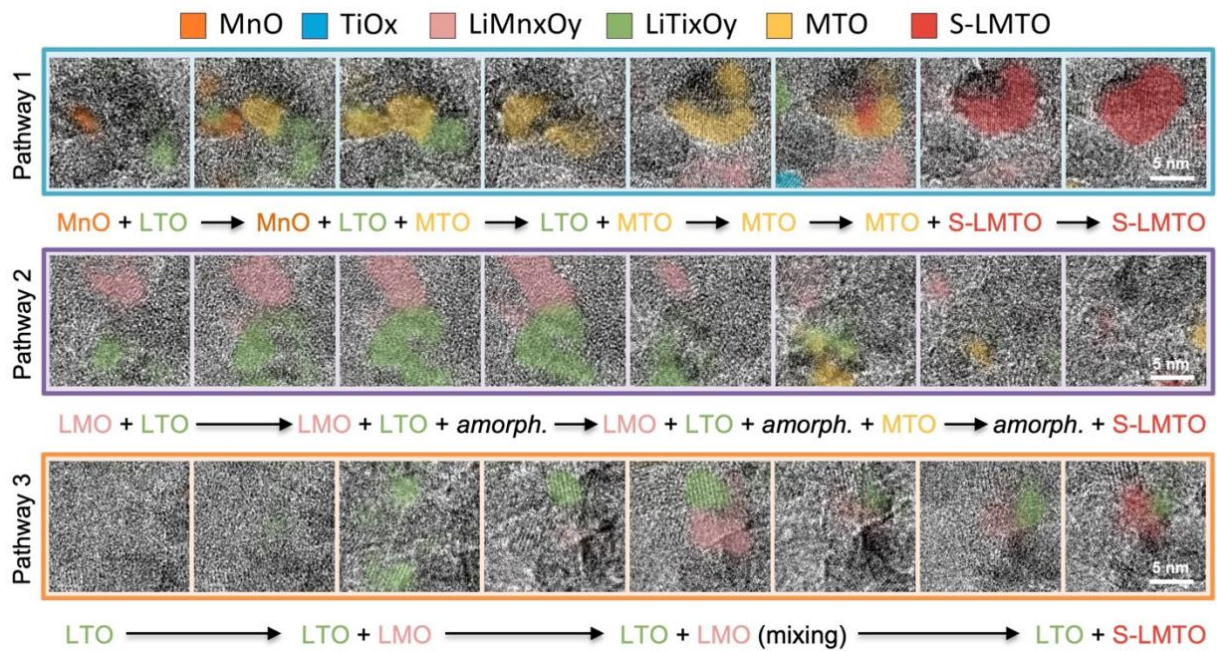

**Figure 3**. **Crystallization pathways captured during the heating sequence till 400 °C**. Each color corresponds to a specific individual phase or a group of phases. Orange color denotes MnO. Blue color denotes the group of titanium oxides, including rutile $TiO_2$, anatase $TiO_2$, and $Ti_2O_3$. Pink color denotes the group of lithium manganese oxides, including $LiMn_2O_4$, $LiMnO_2$, and $Li_2MnO_2$. Green color denotes the group of lithium titanates, including orthorhombic $LiTi_2O_4$, spinel $LiTi_2O_4$, $Li_2Ti_3O_7$, $LiTiO_2$, and $Li_2TiO_3$. Yellow color denotes the group of manganese titanates, including $MnTiO_3$, $Mn_2TiO_4$, $MnTi_2O_4$, and $Mn_2Ti_4O$. Red color denotes S-LMTO. Each unique crystallization pathway is expressed in acronyms underneath each image sequence.

Pathway 1 illustrates a sequential solid-state reaction. Initially segregated nanoscale domains of MnO (orange) and $LiTi_xO_k$ (LTO, green) nucleate and quickly merge to form a transient ternary MTO (yellow) intermediate that continues growing in size. **Table S2** further suggests that the present phase within this MTO group turns out to be a Ti-rich $MnTi_2O_4$ phase. Intriguingly, as heating continues, the $LiMn_xO_k$ (LMO, pink) phase next to the MTO particle gradually amorphizes and potentially provides lithium and manganese source to the MTO phase, which transforms entirely into a single, phase-pure S-LMTO grain (red). Pathway 2 demonstrates a complex amorphization-mediated route. The local environments initially consist of two ternary crystalline intermediates, LMO (pink) and LTO (green). Over an extended period of time, these interfaced ternary phases remain isolated despite their close proximity, likely due to the relatively larger grain sizes and consequently limited diffusion kinetics. At a certain point (the $5^{th}$ panel of pathway 2), the crystalline domains of LMO (pink) and LTO (green) become unstable and dissolve back into an amorphous state. From this local amorphous phase, in a second nucleation event, MTO (yellow) phase appears before transforming into a small S-LMTO grain (weak red). In contrast, pathway 3 begins from a similar initial configuration but evolves along a distinct trajectory. Ternary LTO (green) and LMO (pink) nuclei grow independently in close proximity. The smaller grain sizes (<5 nm) in this case appear to facilitate the subsequent merging and mixing of LTO and LMO phases. Their interfacial mixing successfully generates the S-LMTO target phase (red) and unreacted domain of LTO firmly pinned to the surface of the newly formed S-LMTO crystal. Size effect is obviously present between pathways 2 & 3.

The direct visualization of three different crystallization pathways occurring simultaneously within the same sample underscores the nanoscale heterogeneity in this methanol-based sol-gel synthesis. The FTIR data (**Figure 1b**) suggest that the observed behaviors may arise from heterogeneous metal–acetate coordination within the gel state. The broad envelope at 1560 cm$^{-1}$ is consistent with a reasonably intermixed hybrid gel, potentially shortening diffusion distances, while deconvolution reveals distinct lithium, manganese, and titanium contributions to the acetate band (**Figure S1**). As a result, the gel likely collapses and decomposes gradually and heterogeneously during the 25–400 °C heating stage, giving rise to distinct oxide nanodomains rather than a single uniformly mixed LMTO precursor. Because lithium appears less strongly integrated into the acetate-crosslinked network than manganese and titanium do (**Figure S1**), its redistribution should be more facile during decomposition, favoring the initial formation of compositionally distinct nanoscale regions rather than direct crystallization of fully mixed LMTO. Subsequent phase evolution therefore requires secondary interfacial reaction and lithium redistribution, as observed via *in situ* TEM. Pathway 1 represents the expected thermodynamic trajectory for synthesizing multi-metallic oxides. Because the early-stage binary and ternary oxides nucleate in immediate spatial proximity, rapid solid-state diffusion facilitates structural rearrangement into the expected S-LMTO phase at 400 °C. Conversely, pathways 2 and 3 reveal the nanoscale size effect accompanying localized compositional fluctuations. The intermediate amorphization observed in pathway 2 indicates high local interfacial energy, forcing an energy-intensive dissolution-reprecipitation step at the nanoscale. When ternary intermediate domains like LTO and LMO grow too large before interacting, the extended diffusion lengths and reduced surface-to-volume ratios prevent complete homogenization. Besides size effect, pathway 3 also provides direct empirical proof of the exact origin of macroscopic phase impurities, such as the persistent LTO previously observed in XRD[19].

**Figure 4** extends the *in situ* HRTEM phase mapping to capture the final, high-temperature crystallization stages, specifically detailing the formation of the target DRX-LMTO phase (purple). The image sequences reveal three highly distinct transformation routes. Pathways 1 & 2 are collected from **Movie S3** and pathway 3 is collected from **Movie S4**. Pathway 1 depicts a sequential structural transition that is thermodynamically favored. Here, early intermediates such as LTO (green) and a ternary MTO (yellow) first consolidate into an ordered S-LMTO (red). As thermal energy increases, this highly ordered S-LMTO domain undergoes an order-to-disorder phase transition, progressively transforming into the final DRX-LMTO (purple) phase on the surface of the particle sitting at the top-right corner. Residual LTO is segregated temporarily, then amorphizes and is ultimately incorporated into the DRX lattice.

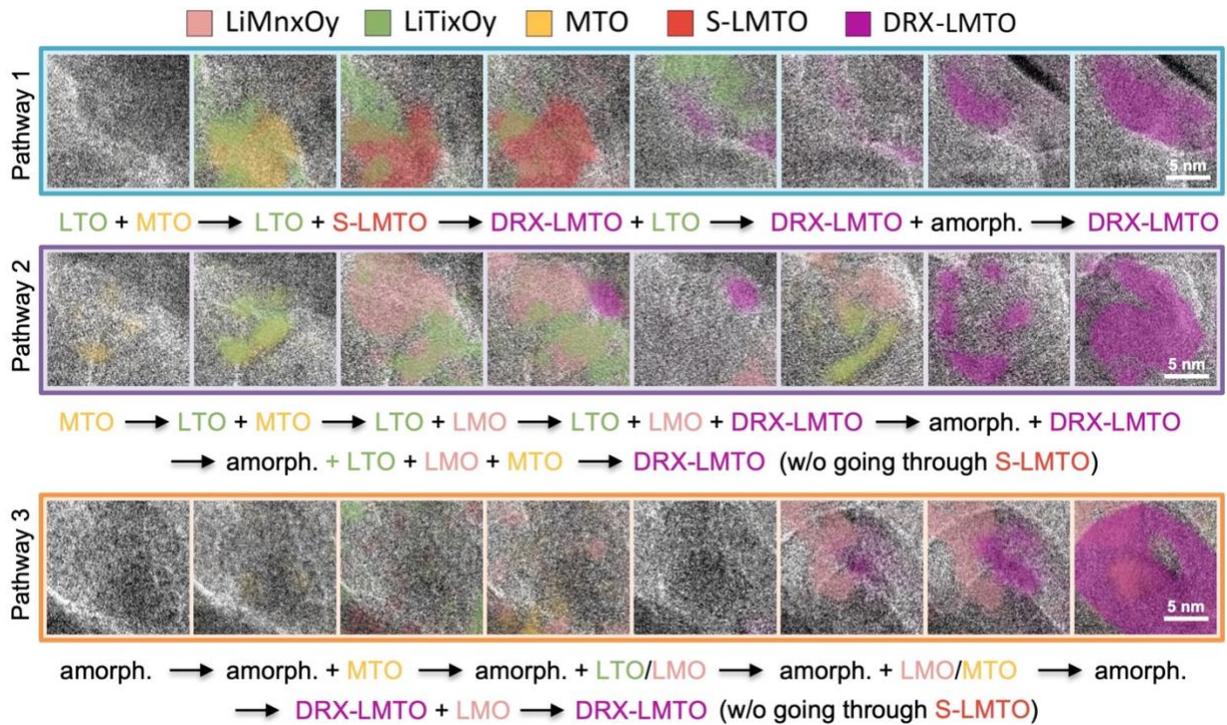

**Figure 4**. **Crystallization pathways identified during the heating sequence from 400 °C to 800 °C**. Each color corresponds to a specific individual phase or a group of phases. Pink color denotes the group of lithium manganese oxides, including $LiMn_2O_4$, $LiMnO_2$, and $Li_2MnO_2$. Green color denotes the group of lithium titanates, including orthorhombic $LiTi_2O_4$, spinel $LiTi_2O_4$, $Li_2Ti_3O_7$, $LiTiO_2$, and $Li_2TiO_3$. Yellow color denotes the group of manganese titanates, including $MnTiO_3$, $Mn_2TiO_4$, $MnTi_2O_4$, and $Mn_2Ti_4O$. Red color denotes S-LMTO and purple color denotes DRX-LMTO. Each unique crystallization pathway is expressed in acronyms underneath each image sequence.

Conversely, pathways 2 and 3 demonstrate non-classical crystallization routes that entirely bypass the S-LMTO intermediate. In pathway 2, distinct domains of MTO, LTO, and LMO (pink) interact in the early stage. DRX-LMTO nucleation then initiates at the interface between LMO and LTO directly. Rather than continue growing into the DRX phase, these domains partially dissolve back into an amorphous state, from which the DRX-LMTO phase nucleates and quickly dominates in the particle of interest. Pathway 3 exhibits a similarly amorphous-dominated trajectory. Transient, highly localized nuclei of MTO, LTO, and LMO emerge from the amorphous matrix but lack sufficient structural stability to coalesce. These nuclei subsequently redissolve into the amorphous phase, which acts as a localized melt, allowing the direct precipitation of a large DRX-LMTO grain that rapidly consumes the remaining LMO fragments.

The *in situ* observations in **Figure 4** reveal the mechanistic complexity underlying the synthesis of the final DRX phase via methanol-based sol-gel process. The results indicate that the formation of the DRX phase does not rely on a single, monolithic reaction mechanism, but is instead highly sensitive to nanoscale fluctuations in local stoichiometry and structural ordering. The system first satisfies local thermodynamic constraints by forming the ordered S-LMTO structure. However, converting this rigid, ordered spinel into a cation-disordered rock salt requires large thermal energy to rearrange the established lattice, explaining why extended, high-temperature calcination is practically required in bulk synthesis. In contrast, pathways 2 and 3 reveal a kinetic shortcut through amorphous phases as intermediates. One hypothesis is that when localized compositional heterogeneity prevents the seamless formation of a stable S-LMTO intermediate, the high local surface energies drive the intermediate nanocrystals to amorphize. This localized amorphization/dissolution effectively resets the structural constraints. Since DRX-LMTO phase is inherently a disordered structure, precipitating it directly from a highly disordered amorphous matrix presents a kinetic advantage over attempting to randomize a rigid spinel lattice. By circumventing the S-LMTO intermediate, these pathways demonstrate that localized amorphization could become a robust, alternate thermodynamic bridge that the sol-gel system utilizes to achieve the final DRX structure. These insights imply the underlying cause for the observed general trend of temperature reduction in sol-gel synthesis of DRX oxides, compared to solid-state synthesis. One can potentially design sol-gel reactions to deliberately induce amorphization pathways over the ones occurring via S-LMTO intermediate.

The TEM results discussed thus far resolve early-stage nanodomain formation and transient intermediate species, whereas XRD provides ensemble information about crystal phase transformations in the bulk. Moving on from a nanoscale perspective, **Figure 5** correlates the ensemble macroscopic phase evolution during high-temperature calcination with its ultimate morphological and electrochemical consequences. **Figure 5a** presents an *in situ* SXRD contour map tracking the structural transitions in the temperature range of 400 - 1000 °C. The evolution of the relative phase fractions was quantified by calculating the phase volume fraction evolution from the SXRD data collected during calcination[19]. As shown in **Figure 5b**, the methanol-derived sample follows a S-LMTO/LTO-mediated transformation pathway toward DRX-LMTO. At the beginning of the analyzed temperature range, substantial fractions of S-LMTO and LTO intermediates are present together with an already significant DRX-LMTO contribution. Upon further heating, the fraction of intermediate phases progressively decreases while the DRX fraction increases continuously. The most pronounced phase conversion occurs between approximately 775 and 880 °C, where both S-LMTO and LTO rapidly diminish in parallel with strong growth of the DRX phase. Above ~880 °C, the sample becomes DRX-dominant, with only minor residual intermediate features persisting at high temperature. These results indicate that the ensemble transformation at later stages proceeds through progressive consumption of S-LMTO and LTO intermediates in favor of DRX phase, while earlier transient features remain more clearly resolved by TEM.

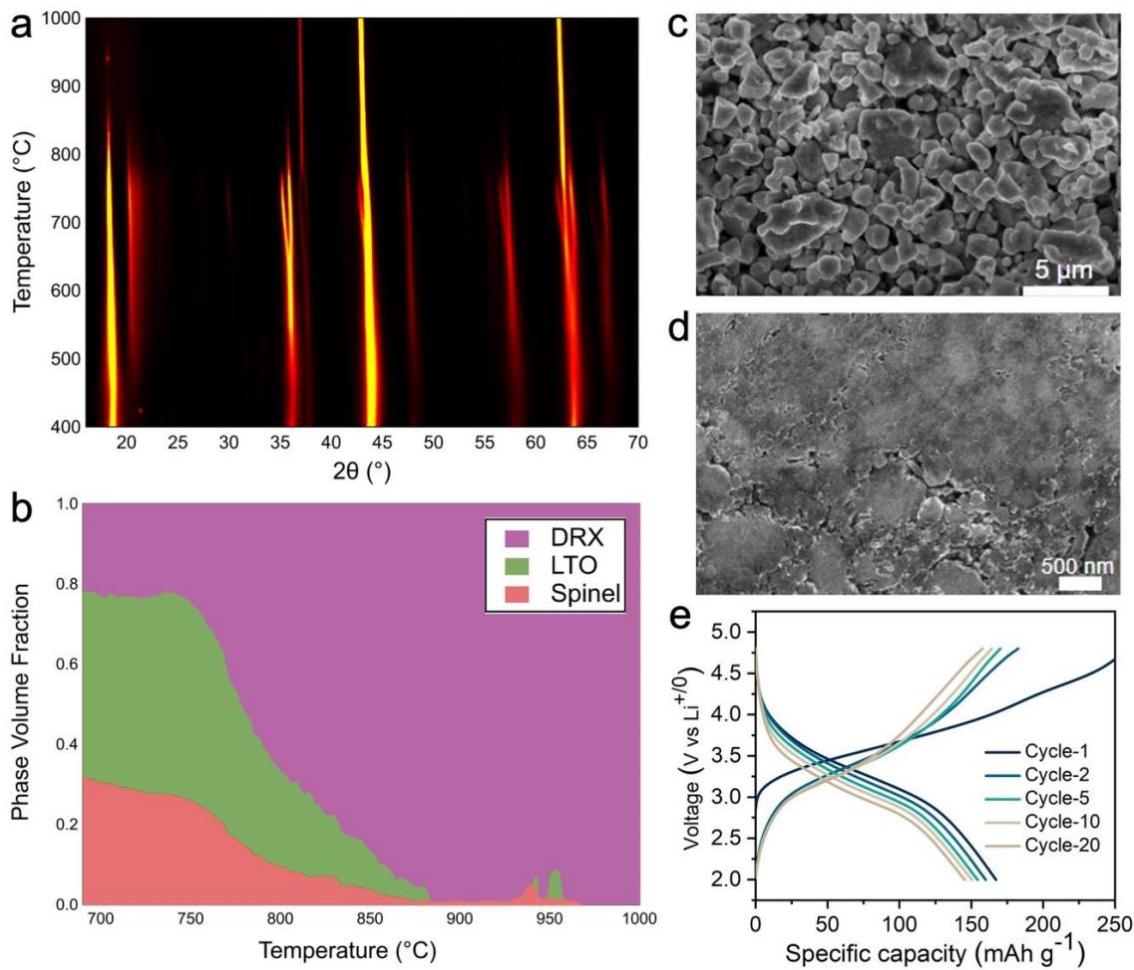

**Figure 5.** *In situ* **synchrotron-based XRD measurements and post-synthesis characterizations of DRX-LMTO samples**. **a**, contour plot showing the evolution of diffraction intensity as a function of temperature and 2θ (converted to Cu Kα for convenience). **b**, quantified evolution of the relative phase volume fractions during heating. The methanol-derived sample exhibits progressive consumption of S-LMTO and LTO intermediates with increasing temperature, accompanied by strong growth of the DRX phase, which becomes dominant above ~880 °C. **c & d**, surface morphologies of DRX-LMTO sample as synthesized powder and casted electrode, respectively. **e**, galvanostatic charge–discharge voltage profiles of the DRX-LMTO material measured over 20 cycles in Li half cells at a current density of 20 mA g$^{-1}$ within a voltage window of 2.0–4.8 V at room temperature. CR2032 coin cells were assembled using composite DRX-LMTO electrodes, lithium metal as the counter/reference electrode, a glass fiber separator, and 1 M LiPF$_6$ in ethylene carbonate/dimethyl carbonate (1:1 v/v) as the electrolyte. Selected voltage profiles from cycles 1, 2, 5, 10, and 20 are shown to illustrate the evolution of the electrochemical response during cycling.

The practical functionality of the sol-gel derived DRX-LMTO material was examined through post-synthesis morphological and electrochemical characterizations. SEM images detailing the morphological appearance of the DRX-LMTO powder after annealing at 1000 °C and electrode casting (after ball-milling) are depicted in **Figure 5c&d**, respectively. This process changes particle morphology and size, as well as compactness. To evaluate the electrochemical performance of the ball-milled DRX-LMTO material in a half cell configuration, Li half cells were assembled and tested by galvanostatic cycling (**Figure 5e**). The material delivers an initial charge capacity of approximately 250 mAh g$^{-1}$ and an initial discharge capacity of about 167 mAh g$^{-1}$, after which the reversible capacity stabilizes near 145 mAh g$^{-1}$ over 20 cycles, consistent with the sloping voltage profile characteristic of a disordered rock salt cathode. Note that the device characteristics were not subject to any optimization and are simply added to demonstrate the potential of such synthesized DRX material.

While *in situ* HRTEM demonstrates the complex nanoscale transformation pathways, *in situ* SXRD shows the dominant routes for DRX formation at macroscopic scale. XRD results (**Figure 5b**) corroborate the three pathways identified in **Figure 4**. The presence of DRX in the early calcination stage around 700 °C can be justified by pathway 3 where DRX precipitation from amorphous precursors is observed. The later dominant route of S-LMTO/LTO transformation to DRX-LMTO aligns with pathway 1 in **Figure 4**. The non-classical transformation pathways observed via *in situ* TEM suggest the possibility to suppress thermodynamically favorable pathways to achieve phase purity of such DRX energy materials at a lower temperature to circumvent such energy-intensive synthesis.

**Conclusions**

In conclusion, this study elucidates the complex crystallization dynamics of DRX-LMTO during thermally activated sol-gel synthesis. By developing a customized, automated virtual dark-field image processing pipeline for *in situ* heating TEM, we overcome traditional analytical limitations in complex chemical systems and directly visualize nanoscale phase transformations. By integrating precisely controlled *in situ* thermal stimuli with this automated phase-mapping algorithm, the approach bridges the observation gap between the wet-chemical gelation state and solid-state calcination. Our results show that the nucleation of sol-gel derived DRX-LMTO is highly sensitive to compositional microenvironments for the chemistry investigated. Crystalline and amorphous formation pathways via different binary and ternary intermediate phases were observed. While some regions follow a classical multi-step transition through S-LMTO or LTO intermediates, others containing intermediate nanocrystals dissolve into a localized amorphous matrix that directly precipitates the DRX structure. Furthermore, *in situ* synchrotron XRD confirms that elevated calcination temperatures (up to 1000 °C) are required to achieve bulk DRX phase purity via S-LMTO/LTO intermediate formation. These findings emphasize that engineering DRX cathodes requires not only balancing macroscopic elemental ratio of precursors, but fundamentally understanding how thermal activation exploits nanoscale structural disorder to achieve final phase purity. Consequently, one could utilize observed nanoscale phase fluctuations to potentially form phase-pure desired DRX structure at a lower synthesis temperature by optimizing precursor chemistries.

**Methods**

*Materials*

LiNO$_3$ (Battery grade, anhydrous), Mn(CH$_3$CO$_2$)$_2$·4H$_2$O (99.99% trace metal basis) titanium (IV) isopropoxide (99.999% trace metal basis), Methanol (anhydrous, 99.8%), and acetic acid solution (suitable for HPLC) were obtained from Sigma Aldrich.

*Sample preparation*

LMTO-DRX was prepared using a sol-gel synthesis method reported previously[19]. In brief, LiNO$_3$ and Mn(CH$_3$CO$_2$)$_2$·4H$_2$O were dissolved in methanol and stirred magnetically at room temperature for 10 min. Titanium(IV) isopropoxide was then introduced dropwise into the continuously stirred solution, followed by an additional 10 min of stirring. Next, 238 µL of acetic acid (3%) and 150 µL of water were added sequentially under continuous stirring at room temperature, serving as the chelating agent and hydrolysis initiator, respectively. To remove the solvent and induce gel formation, the resulting solutions were transferred to a heater-shaker and maintained at 70 °C under constant shaking for 22 hours. After gelation, the gels were dried on a hot plate at 130 °C for 0.5 hours. The sample for the low-temperature *in situ* TEM study was transferred to the TEM at this stage. The remaining samples were pyrolyzed at 400 °C in air on a hot plate and then used for the high-temperature *in situ* heating TEM and in situ heating XRD measurements. For preparation of the cathode active material, the pyrolyzed products were then ground thoroughly using a mortar and pestle, placed into alumina crucibles, and calcined in a tube furnace under flowing argon (200 sccm). The furnace temperature was increased to 1000 °C at a ramp rate of 5 K/min, held for 1 hour, and then allowed to cool naturally.

*Fourier-transform infrared spectroscopy*

FTIR was performed using a Thermo-Fischer Nicolet iS50 instrument in ATR (total reflectance) mode at 25 °C for different solutions. 5 mL of each solution sample was drop-casted and immediately measured. The baseline was collected in the beginning of a set of measurements. The data was collected as absorption and transformed to transmission using the instrument software. All the data were auto-baseline corrected by the software itself.

*Dynamic light scattering*

DLS was performed using a Malvern Zetasizer Nano-ZS with a disposable micro cuvette (40µl). The samples were dispersed in their solvents before gel formation. All measurements were carried out at 25°C and a refraction index (RI) of 1.75 was used for LMTO as an approximation. The viscosity and RI of the suspensions were considered to be the viscosity and RI of the neat solvents (Methanol: viscosity (cP) = 0.5476 and RI = 1.326).

*Scanning/Transmission electron microscopy*

All TEM experiments were conducted using a ThemIS microscope equipped with a Ceta2 CMOS camera and operating at 300 kV. The HAADF image was captured under 300 kV with a probe current of 0.504 pA. The TEM image was taken under 300 kV with a probe current of 1.04 nA. EDS was conducted under 300 kV, using a C2 aperture of 50 mm, a spot size of 7 and probe currents ranging from 0.0117 to 0.9360 nA.

*In situ TEM data collection and pre-processing*

The *in situ* heating experiment was performed using a Gatan 652 single tilt heating holder, capable of heating samples up to 1000 °C. Samples are fixed to the holder using a Tantalum hex-ring, and the holder can accommodate any TEM sample on a 3 mm TEM grid. *In* situ videos were collected under 300 kV with a magnification of 245 kx and an exposure time of 400 ms. The screen current was ~4.04 nA. The dose rate for video collection is ~8000 e$^-$ Å$^{-2}$ s$^{-1}$. Note that thermal heating causes severe sample drifting during video collection, which originates from instrumental limitation. For each of the heating sequences, video collections started when the temperatures reached 400 and 800 °C, respectively, after 15 °C/min temperature ramps, in order to substantially minimize the sample drifting during imaging. Regarding the static, residual sample drifting at final fixed temperatures, intermittent manual stage moving was required to shift the area of interest back to the center of the viewing screen. Collected *in situ* datasets were firstly processed by removing the random, unrecognizable image frames recorded during the intermittent moments for stage moving. The dataset segments were then concatenated and aligned by the linear stack alignment with scale-invariant feature transform (SIFT) technique that is integrated in the ImageJ software. Due to the removals of random, intermittent image frames caused by drifting issues, the lengths of pre-processed *in situ* videos do not necessarily reflect the total time taken to record the video datasets. Nevertheless, the temporal order of the remaining image frames in the *in situ* videos is valid.

*Automated TEM phase mapping and identification*

Automated phase mapping and identification of the *in situ* TEM datasets were performed using a custom Python-based processing pipeline utilizing the SciPy and scikit-image libraries[36,37]. Raw image frames were pre-conditioned using edge apodization (Tukey window) and mean intensity subtraction to eliminate Fourier edge artifacts. High-resolution spot detection in the FFT domain was achieved by applying a Gaussian background subtraction to flatten diffuse scattering, followed by local maxima detection and sub-pixel center-of-mass refinement[33] to extract precise d-spacings and group Friedel conjugate pairs. Diffracting domains were tracked across frames using a nearest-neighbor threshold algorithm and classified against a reference crystallographic library. To resolve phase ambiguity, assignment was determined via a weighted scoring matrix that jointly penalized physical d-spacing error, expected theoretical intensity rank, and temporal deviation from sequential reaction tiers. Spatially resolved phase maps were subsequently generated by applying dynamic Gaussian virtual objective apertures to the localized FFT coordinates, performing an IFFT, and extracting the amplitude fields[34]. Final composite visualizations were rendered by applying spatial Gaussian smoothing to the amplitude blobs, followed by contrast enhancement and alpha blending (50% opacity) to overlay the color-coded phase distributions onto original TEM images.

*In situ synchrotron-based XRD*

Temperature dependent in situ X-ray powder diffraction measurements were carried out at beamline 12.2.2 of the Advanced Light Source at Lawrence Berkeley National Laboratory. The X-ray energy was 23 keV (0.5390 Å), with calibration performed using LaB6 powder (NIST SRM 660). The beam size was approximately 17 μm. Additional details of beamline 12.2.2 have been reported previously[38]. Diffraction data were collected in transmission geometry using a Dectris Pilatus 1M S area detector with a pixel size of 0.172 mm and a 0.45 mm Si sensor. The sample to detector distance was calibrated to 212.813 mm using LaB6 NIST SRM 660. Small amounts of sample, corresponding to about 0.75 mm³ of uncompressed powder, were loaded into sapphire capillaries (80 mm length, 0.75 mm inner diameter, Crytur USA). Sapphire capillaries were used instead of quartz to avoid eutectic melting in the Li containing samples. Temperature and gas environment were controlled with the beamline 12.2.2 lamp heater setup described previously[39,40]. The sample was first heated to 400°C at 25 °C/min under a gas flow of 23.4 Nml/min $N_2$ and 6.6 Nml/min $O_2$, then held at 400°C for 10 minutes while diffraction patterns were continuously recorded with 30 s exposure times. The sample was subsequently heated to 1000 °C at 5°C/min under 30 Nml/min Ar, again with continuous 30 s diffraction collection. At 1000°C, the sample was maintained for 60 minutes under the same Ar flow while 30 s exposures continued. Finally, the sample was rapidly cooled to room temperature at 240°C/min under Ar. Initial processing of the diffraction data, including masking of sapphire reflections and azimuthal integration, was carried out using the open-source Python software Dioptas[40]. Subsequent analysis was performed using a Python workflow based on NumPy[41] and Matplotlib.

*Scanning electron microscopy*

SEM images were collected using a Thermo Fisher Verios 5UC instrument. Images were collected under a beam voltage of 15 kV and beam current of 0.1 nA. Dwell time was 50 ns. Working distance was set to 4.8 mm to optimize the imaging quality.

*Coin cell assembly*

CR2032-type two-electrode coin cells were then fabricated using 10 mm diameter composite working electrodes with an active material loading of about 5 mg cm⁻², lithium metal foil as both counter and reference electrode, and a glass fiber separator (Whatman GF/B). The electrolyte was 1 M LiPF$_6$ (99.99%, Solvionic) dissolved in a 1:1 volume mixture of ethylene carbonate and dimethyl carbonate. For the ball-milled LMTO-DRX electrodes, 280 mg of the as-synthesized active material was combined with 80 mg of Super C65 carbon black (Timcal) and milled for 1 hour in a SPEX800M shaker mill. Electrodes from the ball-milled material were then made by hand mixing 90 wt% of the resulting composite with 10 wt% polytetrafluoroethylene (PTFE, DuPont, Teflon 8A) using a mortar and pestle. The final mixtures were rolled into films approximately 40 μm thick inside an argon-filled glovebox. All cell assembly steps were carried out in a VAC argon glovebox maintained below 1 ppm oxygen and below 1 ppm moisture.

*Electrochemical testing*
The resulting Li half-cells were evaluated by galvanostatic long-term cycling at 20 mA g$^{-1}$ over a voltage window of 2.0–4.8 V at room temperature using a Solartron 1470E testing system.

**Author Contributions**
H.Z., C.M.S.-F., D.C. & T.K. conceived the research concepts. A.H. developed the methanol-based sol-gel synthesis method. A.H., T.K. & M.G. synthesized gel samples and calcined powers. A.H. performed FTIR measurement. R.F.M. performed DLS measurements. A.T.P. performed analysis of FTIR & DLS results. D.C. performed S/TEM measurements and analysis. D.C. developed the automated TEM phase mapping and identification algorithm. T.K. & A.H. performed in situ XRD measurements and T.K. analyzed the data. M.G. collected SEM images. V.S.A. & H.K. assembled coin cells and performed cycling tests. D.C., T.K., A.T.P., H.Z. & C.M.S.-F. co-wrote the manuscript. All authors have participated in the discussion and commented on the manuscript. This work was supervised by C.M.S.-F. & H.Z.

**Acknowledgements**
This work was supported by the U.S. Department of Energy, Office of Science, Office of Basic Energy Sciences (BES), Materials Sciences and Engineering Division under contract no. DE-AC02-05-CH11231 within the D2S2 program (KCD2S2). H.Z. also acknowledges support from the in situ TEM program (KC22ZH). Work at the Molecular Foundry was supported by the Office of Science, Office of Basic Energy Sciences, of the U.S. Department of Energy under Contract No. DE-AC02-05CH11231, under proposal number(s) MFP-10011. Work at the Advanced Light Source (ALS) was done at beamline 12.2.2. The ALS is a DOE Office of Science User Facility under contract no. DE-AC02- 05CH11231. This work was also supported by the Assistant Secretary for Energy Efficiency and Renewable Energy, Transportation Technologies Office

**Data Availability Statement**

All data generated or analyzed in this study are included in the published article and its Supplementary Information files. The raw data and Python codes can be accessed via Zenodo at https://doi.org/10.5281/zenodo.19560989.